\documentclass[twocolumn,showpacs,preprintnumbers,amsmath,amssymb]{revtex4}

\usepackage{graphicx}
\usepackage{dcolumn}
\usepackage{bm}

\begin{document}

\preprint{APS/123-QED}

\title{Observation of thermodynamics originating from a mixed-spin ferromagnetic chain}


\author{H. Yamaguchi$^{1}$, S. C. Furuya$^{2}$, S. Morota$^{1}$, S. Shimono$^{3}$, T. Kawakami$^{4}$, Y. Kusanose$^{5}$, Y. Shimura$^{5}$, K. Nakano$^{1}$, and Y. Hosokoshi$^{1}$}
\affiliation{
$^1$Department of Physical Science, Osaka Metropolitan University, Osaka 599-8531, Japan\\
$^2$Department of Basic Science, University of Tokyo,Tokyo 153-8902, Japan\\
$^3$Department of Materials Science and Engineering, National Defense Academy, Kanagawa 239-8686, Japan\\
$^4$Department of Chemistry, Osaka University, Osaka 560-0043, Japan\\
$^5$Graduate School of Advanced Science and Engineering, Hiroshima University, Hiroshima 739-8530, Japan
}

Second institution and/or address\\
This line break forced

\date{\today}

\begin{abstract}
We present a model compound that forms a mixed-spin ferromagnetic chain.
Our material design, based on the organic radicals, affords a verdazyl-based complex ($p$-Py-V)$_2$$[$Mn(hfac)$_2]$.
The molecular orbital calculations of the compound indicate the formation of a mixed spin-(1/2, 1/2, 5/2) ferromagnetic chain. 
The temperature dependence of magnetic susceptibility reveals its ferromagnetic behavior.
The magnetic specific heat exhibits a double-peak structure and indicates a phase transition at the low-temperature peak. 
The observed characteristics are explained using the quantum Monte Carlo calculations.
Furthermore, the modified spin-wave theory verifies that the double-peak structure of the specific heat significantly reflects the relative ration of the acoustic excitation band and the optical excitation gap.
\end{abstract}

\pacs{75.10.Jm, 
}

\maketitle 
One-dimensional (1D) spin chains have attracted considerable interest for the fundamental understandings of the low-dimensional cooperative phenomena in condensed matter physics, both theoretically and experimentally.
Owing to strong quantum fluctuations, variations in the spin size and exchange interaction result in fascinating quantum many-body phenomena.
Typical examples include the topological ground states found in Haldane chains~\cite{haldane1,haldane2} and quantum magnetization plateau in the trimerized quantum chain~\cite{hase, hida}.
Over the last few decades, mixed-spin chains with two different spins, namely $s$ and $S$, have attracted significant attention as 1D spin systems with the topological quantum properties following the Haldane chain. 
When these two different spins are antiferromagnetically coupled, a ferrimagnetic ground state with a value of $S$-$s$ is generated~\cite{LM}.
Furthermore, the topological argument predicts the appearance of a quantized magnetization plateau~\cite{oshikawa}. 
Our recent studies confirmed the emergence of the abovementioned quantized magnetization plateau in the mixed spin-(1/2, 5/2) Heisenberg chain~\cite{ourmix_1,ourmix_2}.
The thermodynamic properties of such mixed-spin ferrimagnetic chains can be explained by gapless ferromagnetic and gapped antiferromagnetic (AF) spin-wave excitations.

In the case of ferromagnetic coupling, mixed-spin chains are expected to exhibit only ferromagnet properties~\cite{theo1,theo3,theo2,theo4}.
It is presumed that the gapless and the gapped magnon excitations are ferromagnetic and possess acoustic and optical properties, respectively.
The thermodynamic properties of these excitation spectra supposedly appear as two independent energy scales. 
Regarding the specific heat, the appearance of a double-peak structure at $S$ $\textgreater$ $3s$ has been predicted~\cite{theo1}.
However, there are no convincing experimental reports on the thermodynamic properties of mixed-spin ferromagnetic chains. 
Although MnNi(NO$_2$)$_4$(en)$_2$ (en = ethylenediamine) forms a mixed spin-(1, 5/2) ferromagnetic chain and exhibits ferromagnetic contributions in terms of the magnetic susceptibility, the AF interchain couplings in the material induce long-range order at $T_{\rm{N}}$ = 2.45 K~\cite{exp2,exp1,exp4}. 
Furthermore, as the energy scale of the ferromagnetic interaction is relatively close to $T_{\rm{N}}$, the intrinsic low-temperature thermodynamics associated with the excitations are considered to be largely masked by the phase-transition signals.
Although the development of new model compounds is strongly desired, the symmetry and stability of the transition metals makes the formation of mixed-spin ferromagnetic chains difficult.
To the best of our knowledge, no new model compounds have been reported for more than 20 years.


In this letter, we present a model compound combining verdazyl radicals with transition metals~\cite{V_review}, which succeeds in forming a mixed-spin ferromagnetic chain.
We synthesized single crystals of the verdazyl-based complex ($p$-Py-V)$_2$$[$Mn(hfac)$_2]$. 
Our molecular orbital (MO) calculations indicate the formation of a spin-(1/2, 1/2, 5/2) chain composed of two ferromagnetic interactions. 
The observed thermodynamic properties were explained assuming the expected ferromagnetic model by using the quantum Monte Carlo (QMC) calculations.
Furthermore, we confirmed that the double-peak structure of the specific heat significantly reflects the difference between the acoustic excitation band and the optical excitation gap, through the modified spin-wave (MSW) theory.

We prepared $p$-Py-V using the conventional procedure~\cite{gosei} and subsequently synthesized ($p$-Py-V)$_2$$[$Mn(hfac)$_2]$ by following a previously reported procedure for verdazyl-based complexes~\cite{Zn_alt, Mn_alt, Zn_hone, Mn_hone}.
Recrystallization was conducted using acetonitrile yielded dark-green ($p$-Py-V)$_2$$[$Mn(hfac)$_2]$ crystals.
The single crystal diffraction was carried out via the Bruker D8 VENTURE with a PHOTON II detector. 
The magnetic susceptibility was measured using a commercial SQUID magnetometer (MPMS-XL, Quantum Design) down to 1.8 K. 
The experimental result was corrected considering the diamagnetic contributions calculated via Pascal's method. 
The specific heat was measured using a commercial calorimeter (PPMS, Quantum Design) by using a thermal relaxation method down to approximately 0.4 K.
Considering the isotropic nature of organic radical systems, all the experiments were performed using small randomly oriented single crystals.

\begin{figure}[t]
\begin{center}
\includegraphics[width=21pc]{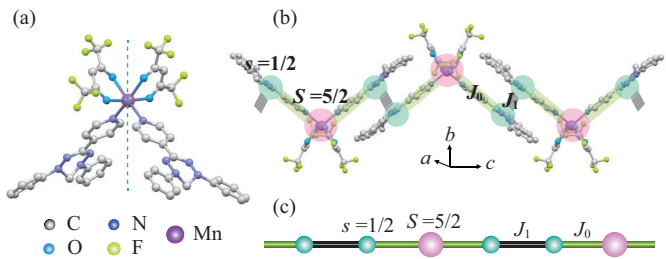}
\caption{(color online) (a) Molecular structure of ($p$-Py-V)$_2$$[$Mn(hfac)$_2]$. Hydrogen atoms are omitted for clarity. The broken line shows the two-fold rotational axis. (b) Crystal structure of ($p$-Py-V)$_2$$[$Mn(hfac)$_2]$ forming 1D spin chain along the $c$-axis. (c) Mixed spin-(1/2,1/2,5/2) ferromagnetic chain composed of $J_{0}$ and $J_{1}$.
}
\end{center}
\end{figure}

The molecular structure of ($p$-Py-V)$_2$$[$Mn(hfac)$_2]$ is shown in Fig. 1(a), where the Mn atom, coordinated by two radicals, yields an octahedral coordination environment~\cite{supple1}.
The verdazyl radical $p$-Py-V and Mn$^{2+}$ have spin-1/2 and 5/2, respectively~\cite{supple1}.
The crystallographic parameters at room-temperature are as follows: monoclinic, space group $C$2/$c$, $a$ =  20.2755(4) $\rm{\AA}$, $b$ = 10.0403(2) $\rm{\AA}$, $c$ = 25.9257(6) $\rm{\AA}$, $\beta$ = 112.1990(10)$^{\circ}$, $V$ = 4886.55(18) $\rm{\AA}^3$, $Z$ = 4, $R$ = 0.0715, and $R_{\rm{w}}$ = 0.1930. 
We performed MO calculations~\cite{MOcal} to evaluate the dominant exchange interactions.
We found ferromagnetic intra- and intermolecular interactions forming a mixed spin-(1/2, 1/2, 5/2) chain~\cite{strecka} along the $c$-axis, as shown in Fig. 1(b) and 1(c).
These interactions are identified as $J_{0}/k_{\rm{B}}$ = $12$ K and $J_{1}/k_{\rm{B}}$ = 24 K ($J_{0}/J_{1}$ = 0.5).
As a two-fold rotational axis parallel to the $b$-axis runs along the center of the molecule, the intramolecular interactions between the radicals and Mn are identical. 
The spin Hamiltonian is given by 
\begin{equation}
\mathcal {H} = -{\sum^{}_{j}}(J_{0}\textbf{{\textit s}}_{3j-2}{\cdot}\textbf{{\textit S}}_{3j-1}+J_{0}\textbf{{\textit S}}_{3j-1}{\cdot}\textbf{{\textit s}}_{3j}+J_{1}\textbf{{\textit s}}_{3j}{\cdot}\textbf{{\textit s}}_{3j+1}), 
\end{equation}
where $\textbf{{\textit s}}$ and $\textbf{{\textit S}}$ are the spin-1/2 and 5/2 operators, respectively.
The nonmagnetic hfac functions as a spacer between the 1D structures; hence, the low dimensionality of the mixed-spin chain is enhanced (see Supplementary Information).
The effective interchain couplings causing the AF order were found to be less than approximately 1/100 of intrachain interactions in absolute values, as described in the following magnetization analysis.

Figure 2(a) shows the temperature dependence of magnetic susceptibility ($\chi$ = $M/H$) and $\chi$$T$ at 0.05 T.
The absence of any significant magnetic field dependence below 0.1 T was confirmed. 
We only observed a monotonic paramagnetic increase in the magnetic susceptibilities with decreasing temperatures.
However, a definite contribution of the ferromagnetic interactions was noted in the $\chi$$T$ values, which increased with the decreasing temperatures.
In the high-temperature region, the value of $\chi$$T$ approached $\sim$ 5.1 emu K/mol, the expected value for the noninteracting spins in the magnetic unit cell.
For mixed-spin ferrimagnetic chains, the minimum of $\chi$$T$ is a typical feature attributed to the AF correlations~\cite{ourmix_1,mix1,mix2,mix3}.
For the model compound presented herein, there is no minimum in $\chi$$T$, indicating that magnetic interactions have only ferromagnetic character.
These findings are consistent with the features indicating the formation of a mixed spin-(1/2,1/2,5/2) ferromagnetic chain.

We examined the $J_{0}/J_{1}$ dependence of $\chi$ for the present model described by Eq. (1) using the QMC method~\cite{QMC}. 
The calculated $\chi$$T$ exhibited a qualitative difference depending on $J_{0}/J_{1}$, while $\chi$ showed a monotonic paramagnetic behavior for all $J_{0}/J_{1}$ values.
Figure 2(b) shows the variation in $\chi$$T$ for selected values of $J_{0}/J_{1}$. 
For each $J_{0}/J_{1}$, the absolute values of the interactions were determined to reproduce the experimental results above 50 K. 
When $J_{0}/J_{1}$ $\gg$ 1, the effective spin model corresponds to a spin-7/2 ferromagnetic chain in the intermediate temperature region where $J_{0}$ correlations are dominant, and $\chi$$T$ approaches $\sim$ 7.9 emu K/mol expected for the spin-7/2 paramagnetic state.
The corresponding $\chi$$T$ behavior is identified in the temperature range between approximately 5 K and 10 K for $J_{0}/J_{1}$ $\textgreater$ 2.5, as shown in Fig. 2(b).
When $J_{0}/J_{1}$ $\ll$ 1, two $s$ = 1/2 spins coupled by the dominant ferromagnetic $J_{1}$ form an effective spin-1; therefore, the system becomes equivalent to a mixed spin-(1, 5/2) ferromagnetic chain.
The calculated results show the monotonic behavior for $J_{0}/J_{1}$ $\ll$ 1 and are consistent with the previous numerical studies~\cite{theo3,theo2}.
To reproduce the experimental results, we considered the effects of the interchain couplings using a mean-field approximation given by $\chi$ = $\chi_{\rm{QMC}}$/${\{}$$1-(zJ_{\rm{inter}}/Ng^{2}{\mu}_{B}^{2})$$\chi_{\rm{QMC}}{\}}$, where $\chi_{\rm{QMC}}$ is the susceptibility obtained through QMC calculations, $z$ is in the number of nearest-neighbor chains, $J_{\rm{inter}}$ is the effective interchain coupling, $N$ is the number of spins, $g$ is the g-factor, and ${\mu}$$_B$ is the Bohr magneton.
Considering the isotropic nature of organic radical and Mn$^{2+}$ ion, we assumed the isotropic $g$-value of 2.0.
We confirmed that the AF interchain couplings suppressed the large increase in the $\chi$$T$ in the low-temperature region and further reproduced the experimental result, as shown in Figs. 2(c)-2(e).
The consistency between the experimental and calculated results was verified using the following parameters: $J_{0}/k_{\rm{B}}$ = 5.6 K, $J_{1}/k_{\rm{B}}$ = 14 K, and $zJ_{\rm{inter}}/k_{\rm{B}}$ = $-0.066$ K for $J_{0}/J_{1}$ = 0.4, as shown in Fig. 2(a).
Although a weak but finite AF $zJ_{\rm{inter}}$ causes an AF long-range order, as shown later in the specific heat, the small $zJ_{\rm{inter}}$ value obtained demonstrates the one-dimensionality of the present system.  
The magnetization curves are also well explained by the mixed-spin ferromagnetic chain using the same parameters~\cite{supple1}. 

\begin{figure}
\begin{center}
\includegraphics[width=20pc]{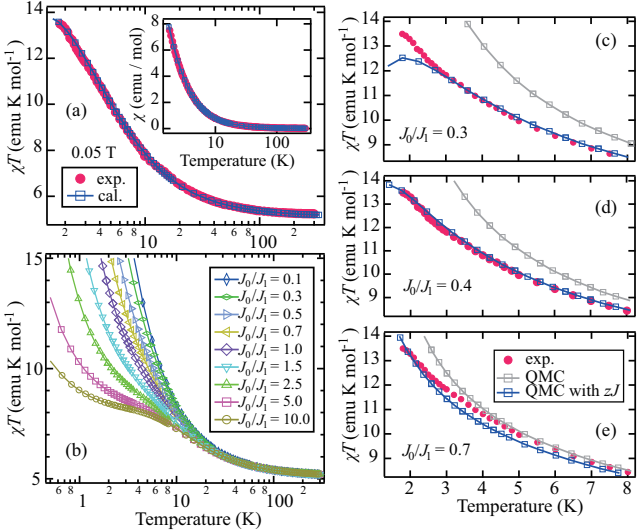}
\caption{(color online) (a) Temperature dependence of magnetic susceptibility ($\chi$ = $M/H$) and $\chi T$ of   ($p$-Py-V)$_2$$[$Mn(hfac)$_2]$ at 0.05 T. 
The solid lines with open squares represent the calculated results for the mixed spin-(1/2,1/2,5/2) ferromagnetic chain with $J_{0}/J_{1}$ = 0.4.
(b) Calculated $\chi T$ of the mixed spin-(1/2,1/2,5/2) ferromagnetic chain for various values of $J_{0}/J_{1}$. 
(c)-(e) Comparison of experimental and calculated results for representative values of $J_{0}/J_{1}$.
The solid lines indicate the QMC calculations and those considering the mean field.
}\label{f2}
\end{center}
\end{figure}

The temperature dependence of the specific heat $C_{\rm{p}}$ is shown in Fig. 3(a). 
We observed a sharp peak at $T_{\rm{N}}$ $=$ 0.95 K, which is indicative of the phase transition to an AF long-range order caused by the finite interchain couplings. 
We evaluated the magnetic specific heat $C_{\rm{m}}$ by subtracting the lattice contribution $C_{\rm{l}}$ and assumed $C_{\rm{l}}$ below 15 K with $C_{\rm{l}}={\alpha}_{1}T^{3}+{\alpha}_{2}T^{5}+{\alpha}_{3}T^{7}$, which has been confirmed to be effective for verdazyl-based compounds~\cite{a235Cl3V, Zn_ferro}.
The constants ${\alpha}_{1}-{\alpha}_{3}$ were determined to reproduce $C_{\rm{m}}$ calculated by the QMC method in the high-temperature region, where $J_{0}/J_{1}$ dependence is almost negligible.
As the result, $C_{\rm{l}}$ with the constants ${\alpha}_{1}=0.045$, ${\alpha}_{2}=-1.6\times10^{-4}$, and ${\alpha}_{3}=2.5\times10^{-7}$ was evaluated.
We calculated $C_{\rm{m}}$ using the QMC method with the same parameters as those obtained from the magnetization analysis for $J_{0}/J_{1}$ =0.3, 0.4, and 0.5.
We found a clear double-peak structure for each of the $J_{0}/J_{1}$ values, and the result for $J_{0}/J_{1}$ = 0.4 well reproduced the experimental behavior including the two peak temperatures, as shown in Fig. 3(b).
Although the experimental $C_{\rm{m}}$ exhibits large values associated with the phase transition accompanied by a large entropy shift at $T_{\rm{N}}$, the consistency of the peak temperatures confirms that the observed thermodynamics originates from the mixed-spin ferromagnetic chain, as expected.
The phase transition signal at the low-temperature peak is associated with the competition between effective interchain couplings and activated gapless magnon excitations, as will be described later. 
The double-peak structure is reduced as $J_{0}/J_{1}$ decreases, which is consistent with the transformation of the effective spin model to a mixed spin-(1, 5/2) ferromagnetic chain without the double-peak structure when $J_{0}/J_{1}$ $\ll$ 1.

\begin{figure}
\begin{center}
\includegraphics[width=21pc]{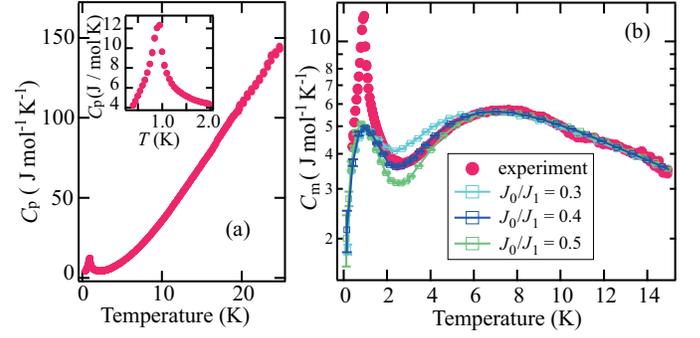}
\caption{(color online) (a) Temperature dependence of the total specific heat $C_{\rm{p}}$ of ($p$-Py-V)$_2$$[$Mn(hfac)$_2]$ at zero field.
The inset shows the extended temperature region associated with the phase transition.
(b) Temperature dependence of the magnetic specific heat $C_{\rm{m}}$.
The solid lines with open squares represent the QMC data for the mixed spin-(1/2,1/2,5/2) forromagnetic chain with $J_{0}/J_{1}$ =0.3, 0.4, and 0.5.
}\label{f3}
\end{center}
\end{figure}

Here, we consider magnon excitations to examine the thermodynamics.  
The three sublattices in Eq. (1) are described using bosonic operators $a_j$-$c_j$, as follows:
\begin{equation*}
s^{z}_{3j-2}=s-a^{\dagger}_{j}a_{j}, s^{-}_{3j-2}=a^{\dagger}_{j}(2s-a^{\dagger}_{j}a_{j})^{1/2}, s^{+}_{3j-2}=(s^{-}_{3j-2})^{\dagger},
\end{equation*}
\begin{equation*}
S^{z}_{3j-1}=S-b^{\dagger}_{j}b_{j}, S^{-}_{3j-1}=b^{\dagger}_{j}(2S-b^{\dagger}_{j}b_{j})^{1/2}, S^{+}_{3j-1}=(S^{-}_{3j-1})^{\dagger}, 
\end{equation*}
\begin{equation}
s^{z}_{3j}=s-c^{\dagger}_{j}c_{j}, s^{-}_{3j}=c^{\dagger}_{j}(2s-c^{\dagger}_{j}c_{j})^{1/2}, s^{+}_{3j}=(s^{-}_{3j})^{\dagger}.
\end{equation}
The Fourier transforms $a_{j}=(3/N){\sum^{}_{k}}e^{-ikj}a_{k}$, $b_{j}=(3/N){\sum^{}_{k}}e^{-ikj}b_{k}$, and $c_{j}=(3/N){\sum^{}_{k}}e^{-ikj}c_{k}$ give the following linear spin-wave Hamiltonian:
\begin{multline}
\mathcal {H} = -\frac{N}{3}(2J_{0}sS+J_{1}s^2)+{\sum^{}_{k}}[(J_{0}S+J_{1}s)(a^{\dagger}_{k}a_{k}+c^{\dagger}_{k}c_{k})\\-J_{1}s{\cos k}(a^{\dagger}_{k}c_{k}+c^{\dagger}_{k}a_{k})+2J_{0}sb^{\dagger}_{k}b_{k}\\-J_{0}\sqrt{sS}{\{}(a^{\dagger}_{k}+c^{\dagger}_{k})b_{k}+b^{\dagger}_{k}(a_{k}+c_{k}){\}}]. 
\label{H_LSW}
\end{multline}
This Hamiltonian is diagonalized through a two-step Bogoliubov transformation~\cite{supple1}, and thus, three types of magnon dispersion relations are obtained, as given by  
\begin{multline*}
{\omega}^{\pm}_{k}=\frac{1}{2}{\{}J_{0}(S+2s)+J_{1}s(1-{\cos k}){\}}\\{\pm}\frac{1}{2}\sqrt{{\{}J_{0}(S-2s)+J_{1}s(1-{\cos k}){\}}^2+8J_{0}^2sS},
\end{multline*}
\begin{equation}
{\omega}^{0}_{k}=J_{0}S+J_{1}s(1+{\cos k}).
\end{equation}
Here, ${\omega}^{-}_{k}$ and (${\omega}^{0}_{k}$, ${\omega}^{+}_{k}$) are the acoustic and optical modes, respectively.
Both excitations feature ferromagnetic characteristics.
Figure 4(a) shows the dispersion relations for $J_{0}/J_{1}$ = 0.4 with the parameters obtained from the magnetization analysis.
The acoustic mode ${\omega}^{-}_{k}$ exhibits ferromagnetic elementary excitation with a quadratic dispersion in the long-wavelength limit.

The linear spin-wave theory \eqref{H_LSW} is effective in deriving the magnon dispersions but ineffective in predicting thermodynamic quantities at finite temperatures.
To overcome this difficulty, we employed the MSW theory~\cite{MSW1,MSW2} and evaluated the specific heat attributed to the magnon excitations.
The MSW theory avoids the thermal divergence of the number of bosons to describe the thermodynamics.
Following Ref.~\cite{MSW1}, we constrained the total magnetization to be zero.
In terms of magnons, this constraint can be written as   
\begin{equation}
S+2s-\frac{3}{N}{\sum^{}_{k}}({n}^{+}_{k}+{n}^{0}_{k}+{n}^{-}_{k})=0, 
\label{constraint}
\end{equation}
where ${n}^{\lambda}_{k}=1/(e^{({\omega}^{\lambda}_{k}+\mu)/k_B T}-1)$.
We added a chemical potential $\mu$ to the dispersion $\omega_k^\lambda$.
We determine the chemical potential $\mu$ self-consistently within the framework of the MSW theory.
The free energy can be represented as
\begin{equation}
F=E-T{\mathcal{S}}+{\mu'}{\{}S+2s-\frac{3}{N}{\sum^{}_{k}}({n}^{+}_{k}+{n}^{0}_{k}+{n}^{-}_{k}){\}}, 
\end{equation}
where $E$ and ${\mathcal{S}}$ are the internal energy and entropy, respectively~\cite{supple1}.
The constraint given in Eq. \eqref{constraint} was included in the free energy with the Lagrange multiplier $\mu'$.
We determined the parameters by minimizing the free energy and calculated specific heat, as shown in Fig. 4(b).
In particular, we obtain $\mu'=\mu$.
A clear consistency with the QMC results, as seen for the double-peak structure and its $J_{0}/J_{1}$ dependence, was proven. 
As the current procedure does not include sufficient intermagnon interactions, certain quantitative differences was noted, and the broad peak appeared at a slightly higher temperature than that for the QMC results, as shown in Fig. 4(c).
However, it should be noted that the position of the low-temperature peak was consistent in both the the calculated and experimental results.
A previous spin-wave study on the mixed ferromagnetic chain predicted that the low-temperature peak appears when $W^{-}/{\Delta}$ ${\ll}$ 1~\cite{theo1}, where $W^{-}$ and $\Delta$ correspond to the acoustic excitation band ${\omega}^{-}_{k}(\pi)$ and the optical excitation gap ${\omega}^{0}_{k}(\pi)$, respectively.
In this study, as $W^{-}/{\Delta}$ was evaluated to be approximately 0.18, sufficient separation of the energy branch occurred, yielding a double-peak structure with a clear low-temperature peak.
On heating from the lowest $T$, near the low-temperature peak the gapless magnon excitations become active, compete with the weak interchain couplings, and finally lead to a phase transition to the disordered state, as seen in the experiment.

\begin{figure}
\begin{center}
\includegraphics[width=20pc]{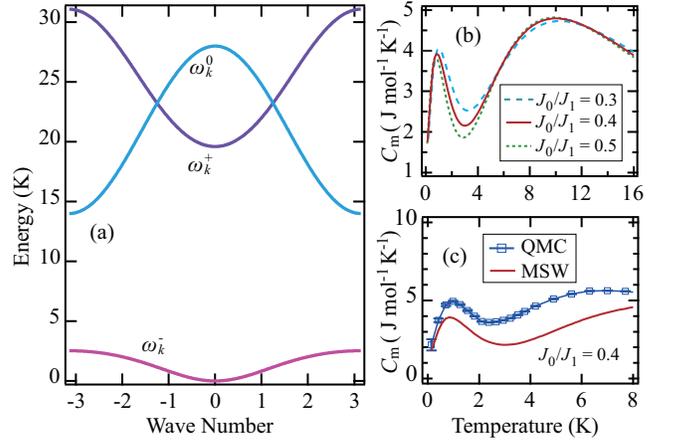}
\caption{(color online) (a) Single-magnon excitation spectra of the mixed spin-(1/2,1/2,5/2) ferromagnetic chain for $J_{0}/J_{1}$ = 0.4. ${\omega}^{-}_{k}$ and (${\omega}^{0}_{k}$, ${\omega}^{+}_{k}$) indicate acoustic and optical modes, respectively. (b) Magnetic specific heat $C_{\rm{m}}$ calculated by the MSW theory for $J_{0}/J_{1}$ =0.3, 0.4, and 0.5. (c) Low-temperature region of $C_{\rm{m}}$ for the chain model with $J_{0}/J_{1}$ = 0.4 obtained from QMC and MSW calculations. 
}\label{f4}
\end{center}
\end{figure}

In summary, we succeeded in synthesizing single crystals of the verdazyl-based complex ($p$-Py-V)$_2$$[$Mn(hfac)$_2]$.
MO calculations indicated that the intramolecular interaction $J_0$ and the intermolecular interaction $J_1$ form a mixed spin-(1/2, 1/2, 5/2) ferromagnetic chain. 
We examined the dependence of magnetic susceptibility on the $J_0$/$J_1$ values through QMC calculations and explained the observed ferromagnetic behavior using the parameters $J_{0}/k_{\rm{B}}$ = 5.6 K and  $J_{1}/k_{\rm{B}}$ = 14 K ($J_{0}/J_{1}$ = 0.4).
We also demonstrated the consistency in the peak positions of the double-peak structure of the specific heat using the QMC calculation, while the experimental values include the contribution of the phase transition due to the weak but finite interchain interactions.
Furthermore, we considered magnon excitations to examine the thermodynamics, and the calculations using the MSW theory described the specific heat with the double-peak structure. 
This work provides new insight into the thermodynamics of mixed-spin ferromagnetic chains, and will thereby stimulate further studies on the thermodynamic properties associated with 1D mixed-spin systems.

\begin{acknowledgments}
We thank Y. Iwasaki and Y. Kono for valuable discussions.
S.C.F. was supported by JSPS Grants-in-Aid for Transformative Research Areas (A) ``Extreme Universe'' (Nos. JP21H05191 and 21H05182) and JSPS KAKENHI Grant Nos. JP20K03769 and JP21K03465.
\end{acknowledgments}



\begin{thebibliography}{99}
\bibitem{MOcal}
The MO calculations were performed using the UB3LYP method in the Gaussian 09 program package. 
The basis sets are 6-31G (intermolecule) and 6-31G($d$,$p$) (intramolecule). 
For the estimation of intermolecular magnetic interaction, we applied the previously reported evaluation scheme~\cite{MO_method}.
\bibitem{QMC}
Quantum Monte Carlo (QMC) calculations were performed with a system size of 96 under periodic boundary conditions using using ALPS~\cite{QMC_ref1,QMC_ref2}. 
QMC code is based on the directed loop algorithm in the stochastic series expansion representation~\cite{QMC_ref3}.
\end{thebibliography}
\end{document}